\documentclass[11pt]{article}
\usepackage{mathptm}
\usepackage{graphicx}
\usepackage{cite}
\usepackage{algorithm}
\usepackage{fullpage}
\usepackage{algorithmic}

\newif\ifFull
\Fullfalse

\ifFull
\else
\renewcommand{\subsection}[1]{\paragraph{#1.}}
\fi

\newtheorem{lemma}{Lemma}
\newtheorem{theorem}[lemma]{Theorem}

\newenvironment{proof}{\noindent{\bf Proof:}}{\hspace*{\fill}\rule{6pt}{6pt}\smallskip}
\newcommand{\R}{{\mbox{\textbf{R}}}}

\title{The Skip Quadtree: A Simple Dynamic \\
    Data Structure for Multidimensional Data}

\author{David Eppstein\footnotemark[2] \and
Michael T. Goodrich\footnotemark[2] \and Jonathan Z. Sun\footnotemark[2]}

\date{}

\begin{document}
\maketitle
\renewcommand{\thefootnote}{\fnsymbol{footnote}}

\footnotetext[2]{Dept.~of Computer Science,
University of California, Irvine, CA 92697-3425, USA.
\texttt{\{eppstein,goodrich,zhengsun\}(at)ics.uci.edu}.}



\renewcommand{\thefootnote}{\arabic{footnote}}

\begin{abstract}
We present a new multi-dimensional data structure, which we call
the skip quadtree (for point data in $\R^2$) or the skip octree (for
point data in $\R^d$, with constant $d>2$).
Our data structure combines the best features of two
well-known data structures, in that it has the well-defined
``box''-shaped regions of region quadtrees and the
logarithmic-height search and update hierarchical structure of skip lists.
Indeed, the bottom level of our structure is exactly a region quadtree
(or octree for higher dimensional data).
We describe efficient algorithms for inserting
and deleting points in a skip quadtree, as well as fast methods for
performing point location and approximate range queries.
\end{abstract}

\section{Introduction}
Data structures for multidimensional point data are of
significant interest in the computational geometry, computer
graphics, and scientific data visualization literatures.
They allow point data to be stored and searched efficiently, for
example to perform range queries to report (possibly approximately)
the points that are contained in a given query region.
We are interested in this paper in data structures for multidimensional
point sets that are dynamic, in that they allow for fast point
insertion and deletion, as well as efficient, in that they use linear
space and allow for fast query times.

\subsection{Related Previous Work}
Linear-space multidimensional data structures typically
are defined by hierarchical
subdivisions of space, which give rise to tree-based
search structures.
That is, a hierarchy is defined by associating with
each node $v$ in a tree $T$
a region $R(v)$ in $\R^d$
such that the children of $v$ are associated with subregions of
$R(v)$ defined by some kind of ``cutting'' action on $R(v)$.
Examples include:
\begin{itemize}
\item
\emph{quadtrees}~\cite{o-mtuas-82}:
regions are defined by squares in the plane, which are
subdivided into four equal-sized squares for any regions
containing more than a single point. So each internal
node in the underlying tree has four children and regions have
optimal aspect ratios (which is useful for many types of queries).
Unfortunately, the tree can have arbitrary depth, independent even of
the number of input points.  Even so, point insertion and deletion is
fairly simple.
\item
\emph{octrees}~\cite{ftyk-oasm-83,o-mtuas-82}:
regions are defined by hypercubes in $\R^d$, which are
subdivided into $2^d$ equal-sized hypercubes for any regions
containing more than a single point. So each internal
node in the underlying tree has $2^d$ children and, like quadtrees,
regions have optimal aspect ratios and point insertion/deletion is
simple, but the tree can have arbitrary depth.
\item
\emph{$k$-d trees}~\cite{b-mbstu-75}:
regions are defined by hyperrectangles in $\R^d$,
which are subdivided into two hyperrectangles using an
axis-perpendicular cutting hyperplane through the median point,
for any regions containing more than two points.
So the underlying tree is binary and has $\lceil\log n\rceil$ depth.
Unfortunately, the regions can have arbitrarily large aspect ratios,
which can adversely affect the efficiencies of some queries.
In addition,
maintaining an efficient $k$-d tree subject to point insertions and
removal is non-trivial.
\item
\emph{compressed quad/octrees}~\cite{as-dchan-99,b-acpqh-93,bet-pcqqt-93,c-faann-83}:
regions are defined in the same way
as in a quadtree or octree (depending on the dimensionality), but
paths in the tree consisting of nodes with only one non-empty child
are compressed to single edges.  This compression allows regions to
still be hypercubes (with optimal aspect ratio), but it changes the
subdivision process from a four-way cut to a reduction to at most four
disjoint hypercubes inside the region.  It also forces the height of
the (compressed) quad/octree to be at most $O(n)$.
This height bound is still not very efficient, of course.
\item
\emph{balanced box decomposition (BBD) trees}
\cite{am-annqf-93,am-ars-00,amnsw-oaann-98}:
regions are defined by hypercubes
with smaller hypercubes subtracted away, so that the height of the
decomposition tree is $O(\log n)$.  These regions have good aspect
ratios, that is, they are ``fat''~\cite{ekns-ddsfo-00,ers-ufwsc-93b},
but they are not convex, which limits some of the
applications of this structure.
In addition, making this structure dynamic appears non-trivial.
\item
\emph{balanced aspect-ratio (BAR) trees}~\cite{d-bart-99,dgk-bartc-01}:
regions are defined by
convex polytopes of bounded aspect ratio, which are subdivided
by hyperplanes perpendicular to one of a set of $2d$ ``spread-out''
vectors so that the height of the decomposition tree is $O(\log n)$.
This structure has the advantage of having convex regions and
logarithmic depth, but the regions are no longer hyperrectangles (or
even hyperrectangles with hyperrectangular ``holes'').
In addition, making this structure dynamic appears non-trivial.
\end{itemize}
This summary is, of course, not a complete review of existing work on
space partitioning
data structures for multidimensional point sets.
The reader interested in further study of these topics is encouraged
to read the book chapters by
Asano \textit{et al.}~\cite{aeiim-pubtc-85},
Samet~\cite{s-sds-95,s-msd-99,s-msds-05},
Lee~\cite{l-isrps-05}, Aluru~\cite{a-qo-05},
Naylor~\cite{n-bspt-05},
Nievergelt and Widmayer~\cite{nw-sdscd-00},
Leutenegger and Lopez~\cite{ll-rt-05},
Duncan and Goodrich~\cite{dg-agqs-05}, and Arya and
Mount~\cite{am-cgpl-05}, as well as the books by de~Berg
\textit{et al.}~\cite{bkos-cgaa-00} and Samet~\cite{s-sdsqo-89,s-asdsc-90}.

\subsection{Our Results}
In this paper we present a dynamic data structure for
multidimensional data, which we call the \emph{skip quadtree} (for
point data in $\R^2$)
or the \emph{skip octree} (for point data in $\R^d$, for fixed $d>2$).
For the sake of simplicity, however, we will often use the term
``quadtree'' to refer to both the two- and multi-dimensional
structures.
This structure provides a hierarchical view of a quadtree in a
fashion reminiscent of the way the skip-list
data structure~\cite{p-slpab-90,mps-dsl-92}
provides a hierarchical view of a linked list.
Our approach differs fundamentally from previous techniques for
applying skip-list hierarchies to
multidimensional point data~\cite{ln-ahsrs-02,n-sldsm-94} or
interval data~\cite{h-islds-91}, however, in that the bottom-level structure
in our hierarchy is not a list---it is a tree.
Indeed, the bottom-level structure
in our hierarchy is just a compressed
quadtree~\cite{as-dchan-99,b-acpqh-93,bet-pcqqt-93,c-faann-83}.
Thus, any operation that can be performed with a quadtree can be
performed with a skip quadtree.
More interestingly, however, we show that point location and
approximate range queries can be performed in a skip quadtree
in $O(\log n)$ and $O(\epsilon^{1-d}\log n+k)$ time,
respectively, where $k$ is the
size of the output in the approximate range query case, for constant
$\epsilon>0$.
We also show that point insertion and deletion can be performed
in $O(\log n)$ time.
We describe both randomized and deterministic versions of our data
structure, with the above time bounds being expected bounds for the
randomized version and worst-case bounds for the deterministic
version.

Due to the balanced aspect ratio of their cells, quadtrees have many
geometric applications including range searching, proximity problems,
construction of well separated pair decompositions, and quality
triangulation.  However, due to their potentially high depth,
maintaining quadtrees directly can be expensive.  Our skip quadtree
data structure provides the benefits of quadtrees together with fast
update and query times even in the presence of deep tree branches, and
is, to our knowledge, the first balanced aspect ratio subdivision with
such efficient update and query times.  We believe that this data
structure will be useful for many of the same applications as
quadtrees. In this paper we demonstrate the skip quadtree's benefits
for two simple types of queries: point location within the quadtree
itself, and approximate range searching.

\section{Preliminaries}

In this section we discuss some preliminary conventions
we use in this paper.

\subsection{Notational Conventions}
Throughout this paper we use
$Q$ for a quadtree and $p,q,r$ for squares or quarters of squares
associated with the nodes of $Q$.
We use
$S$ to denote a set of points and $x,y,z$ for points in $\R^d$.
We let $p(x)$ denote the
smallest square in $Q$ that covers the location of some point $x$, regardless
if $x$ is in the underlying point set for $Q$ or not.
Constant $d$ is reserved for the dimensionality of our search space,
$\R^d$, and we assume throughout that $d\ge 2$ is a constant.
In $d$-dimensional space we still use the term
``square'' to refer to a $d$-dimensional
cube and we use ``quarter'' for any of the $1/2^d$ partitions of a square
$r$ into squares having the center of $r$ as a corner and sharing
part of $r$'s boundary.
A square $r$ is identified by its center
$c(r)$ and its half side length $s(r)$.

\subsection{The Computational Model}

As is standard practice
in computational geometry algorithms dealing
with quadtrees and octrees
(e.g., see~\cite{bet-pcqqt-93}), we assume in this paper that certain
operations on points in $\R^d$ can be done in constant time.
In real applications, these operations are typically performed
using hardware operations that have running times similar to operations
used to compute linear intersections and perform point/line comparisons.
Specifically, in arithmetic terms, the computations needed to perform point
location in a quadtree, as well as update and range query operations,
involve finding the
most significant binary digit at which two coordinates of two
points differ.
This can be done in $O(1)$ machine instructions if
we have a most-significant-bit instruction, or by using
floating-point or extended-precision normalization.
If the coordinates are not in binary fixed or floating point, such
operations may also involve computing integer floor and ceiling
functions.

\subsection{The Compressed Quadtree}\label{compressed}
As the bottom-level structure in a skip quadtree is a compressed
quadtree~\cite{as-dchan-99,b-acpqh-93,bet-pcqqt-93,c-faann-83},
let us briefly review this structure.

The compressed quadtree is defined in terms of an underlying
(standard) quadtree for the same point set; hence, we define the compressed
quadtree by identifying which squares from the standard quadtree
should also be included in the compressed quadtree.
Without loss of generality, we can assume that the
center of the root square (containing the entire point set of
interest) is the origin and the half side length
for any square in the quadtree is a power of $2$. A point $x$ is
contained in a square $p$ iff $-s(p)\leq x_i-c(p)_i<s(p)$ for each
dimension $i\in [1,\cdots,d]$. According to whether $x_i-c(p)_i<0$
or $\geq 0$ for all dimensions we also know in which quarter of
$p$ that $x$ is contained.

\begin{figure}[htb]
\vspace*{-10pt}
\begin{center}
\includegraphics[scale=0.6]{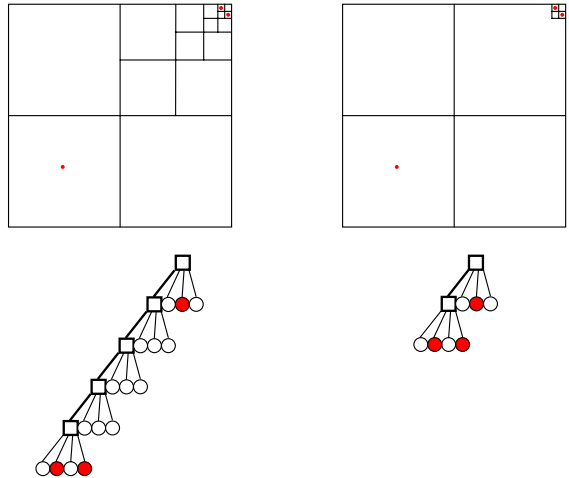}
\end{center}
\vspace*{-15pt} \caption{A quadtree containing 3 points (left) and
its compressed quadtree (right). Below them are the pointer
representations, where a square or an interesting square is
represented by a square, a point by a solid circle and an empty
quarter by a hollow circle. The 4 children of each square are
ordered from left to right according to the I, II, III, IV
quadrants.} \label{Fig-skip-0}
\end{figure}

Define an \emph{interesting square} of a (standard) quadtree to be one that
is either the root of the quadtree or that has two or more nonempty
children.
Then it is clear that any quadtree square $p$ containing
two or more points contains a unique largest interesting square
$q$ (which is either $p$ itself of a descendent square of $p$ in the
standard quadtree).
In particular, if $q$ is the largest interesting square for $p$,
then $q$ is the LCA in the quadtree of the points contained in
$p$.
We compress the (standard) quadtree to explicitly store only the
interesting squares, by splicing out the non-interesting squares
and deleting their empty children from the original quadtree. That
is, for each interesting square $p$, we store $2^d$ bi-directed
pointers one for each $d$-dimensional quarter of $p$. If the quarter contains
two or more points, the pointer goes to the largest interesting
square inside that quarter; if the quarter contains one point, the
pointer goes to that point; and if the quarter is empty, the
pointer is NULL.
We call this structure a
\emph{compressed quadtree}~\cite{as-dchan-99,b-acpqh-93,bet-pcqqt-93,c-faann-83}.
(See Fig.~\ref{Fig-skip-0}.)

A compressed $d$-dimensional quadtree $Q$ for $n$ points has size
$O(n)$, but its worst-case height is $O(n)$,
which is inefficient yet nevertheless
improves the arbitrarily-bad worst-case height of a standard
quadtree.
These bounds follow immediately from the fact
that there are $O(n)$ interesting squares, each of which has size $O(2^d)$.

With respect to the arithmetic operations needed when dealing with
compressed quadtrees,
we assume that we can do the following operations in $O(1)$ time:

\begin{itemize}
    \item Given a point $x$ and a square $p$,
    decide if $x\in p$ and if yes, which
    quarter of $p$ contains $x$.
    \item Given a quarter of
    a square $p$ containing two points $x$ and $y$,
    find the largest interesting square inside this quarter.
    \item Given a quarter of $p$ containing an interesting square
    $r$ and a point $x\not\in r$, find the largest interesting
    square inside this quarter.
\end{itemize}

A standard search in a compressed quadtree $Q$ is to locate the
quadtree square containing a given point $x$.
Such a search starts from the quadtree
root and follows the parent-child pointers, and returns the
smallest interesting square $p(x)$ in $Q$ that covers the location
of $x$. Note that $p(x)$ is either a leaf node of $Q$ or an
internal node with none of its child nodes covering the location of
$x$. If the quarter of $p(x)$ covering the location of $x$
contains exact one point and it matches $x$, then we find $x$ in
$Q$. Otherwise $x$ is not in $Q$, but the smallest interesting
square in $Q$ covering the location of $x$ is found.
The search proceeds in a top-down fashion from the root, taking
$O(1)$ time per level; hence,
the search time is $O(n)$.

Inserting a new point starts by locating the interesting square $p(x)$
covering $x$.
Inserting $x$ into an empty quarter of $p(x)$ only takes $O(1)$
pointer changes. If the quarter of $p(x)$ $x$ to be inserted into
already contains a point $y$ or an interesting square $r$, we
insert into $Q$ a new interesting square $q\subset p$ that
contains both $x$ and $y$ (or $r$) but separates $x$ and $y$ (or
$r$) into different quarters of $q$. This can be done in $O(1)$
time. So the insertion time is $O(1)$, given $p(x)$.

Deleting $x$ may cause its covering interesting square $p(x)$ to
no longer be interesting. If this happens, we splice $p(x)$ out
and delete its empty children from $Q$. Note that the parent node
of $p(x)$ is still interesting, since deleting $x$ doesn't change
the number of nonempty quarters of the parent of $p(x)$.
Therefore, by splicing out at most one node (with $O(1)$ pointer
changes), the compressed quadtree is updated correctly. So a
deletion also takes $O(1)$ time, given $p(x)$.

{\theorem Point-location searching, as well as point insertion and
deletion, in a compressed $d$-dimensional quadtree of $n$ points
can be done in $O(n)$ time.}

Thus, the worst-case time for querying a compressed quadtree
is no better than that of brute-force searching of an unordered set
of points.
Still, like a standard quadtree, a compressed quadtree is unique given a set
of $n$ points and a (root) bounding box, and this uniqueness allows
for constant-time update operations if we have already identified the
interesting square involved in the update.
Therefore, if we could find a faster way to query a compressed
quadtree while still allowing for fast updates, we could construct an
efficient dynamic multidimensional data structure.

\section{The Randomized Skip Quadtree}\label{randomized}
In this section, we describe and analyze the randomized skip quadtree
data structure, which provides a hierarchical view of a compressed
quadtree so as to allow for logarithmic expected-time querying and
updating, while keeping the expected space bound linear.

\subsection{Randomized Skip Quadtree Definition}

The randomized skip quadtree is defined
by a sequence of compressed quadtrees that are respectively defined on
a sequence of subsets of the input set $S$.
In particular,
we maintain a sequence of subsets of the input points $S$,
such that $S_0=S$, and, for $i>0$, $S_i$ is sampled from $S_{i-1}$ by
keeping each point with probability 1/2.
(So, with high probability, $S_{2\log n} = \emptyset$.)
For each $S_i$, we form a (unique)
compressed quadtree $Q_i$ for the points in $S_i$.
We therefore view the $Q_i$'s as forming a sequence of levels
in the skip quadtree, such that $S_0$ is the bottom level (with its
compressed quadtree defined for the entire set $S$)
and $S_{top}$ being the top level, defined as the lowest level with
an empty underlying set of points.

\begin{figure}[hbt]
\vspace*{-10pt}
\begin{center}
\includegraphics[scale=0.6]{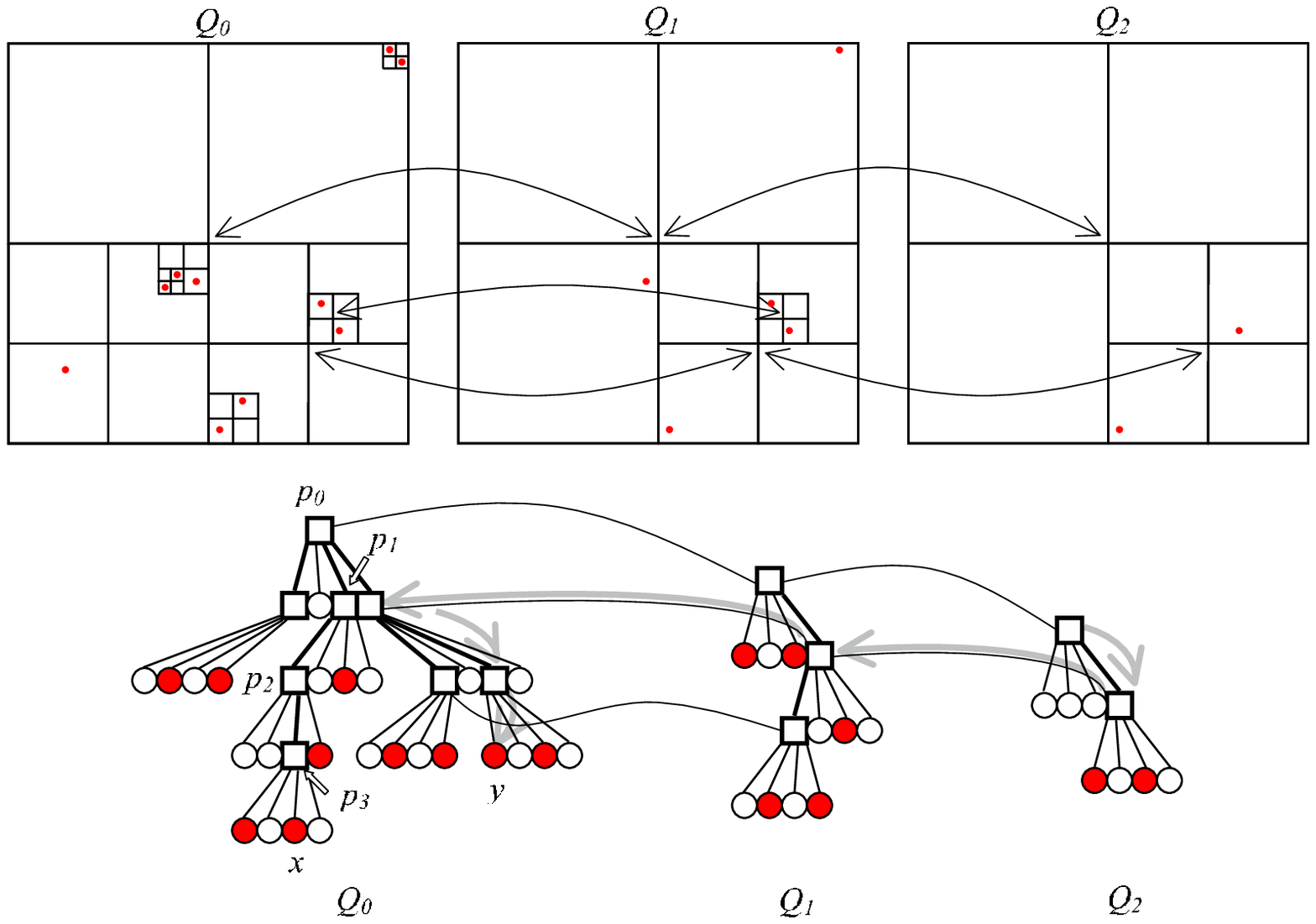}
\end{center}
\vspace*{-10pt} \caption{A randomized skip quadtree consists of
$Q_0$, $Q_1$ and $Q_2$. (Identical interesting squares in two
adjacent compressed quadtrees are linked by a double-head arrow
between the square centers.)}\label{Fig-skip-1}
\end{figure}

Note that if a square $p$ is
interesting in $Q_i$, then it is also interesting in $Q_{i-1}$.
Indeed, this \emph{coherence} property between levels in the skip
quadtree is what facilitates fast searching.
For each
interesting square $p$ in a compressed quadtree $Q_i$, we add two
pointers: one to the same square $p$ in $Q_{i-1}$ and another to
the same square $p$ in $Q_{i+1}$ if $p$ exists in $Q_{i+1}$, or
NULL otherwise.
The sequence of $Q_i$'s and $S_i$'s, together with these auxiliary
pointers define the skip quadtree.
(See Fig.~\ref{Fig-skip-1}.)

\subsection{Search, Insertion, and Deletion in a Randomized Skip Quadtree}

To find the smallest square in $Q_0$ covering the location of a
query point $x$, we start with the root square $p_{l,start}$ in
$Q_l$. ($l$ is the largest value for which $S_l$ is nonempty and
$l$ is $O(\log n)$ w.h.p.) Then we search $x$ in
$Q_l$ as described in Section~\ref{compressed}, following the
parent-child pointers until we stop at the minimum interesting
square $p_{l,end}$ in $Q_l$ that covers the location of $x$. After
we stop searching in each $Q_i$ we go to the copy of $p_{i,end}$
in $Q_{i-1}$ and let $p_{i-1,start}=p_{i,end}$ to continue
searching in $Q_{i-1}$. (See the searching path of $y$ in
Fig.~\ref{Fig-skip-0}.)

{\lemma For any point $x$, the expected number of searching steps
within any individual $Q_i$ is constant. \label{25}}

\begin{proof}
Suppose the searching path of $x$ in $Q_i$ from the root of $Q_i$
is $p_0,p_1,\cdots, p_m$. (See $Q_0$ in Fig.~\ref{Fig-skip-1}.)
Consider the probability $Pr(j)$ of $Event(j)$ such that $p_{m-j}$
is the last one in $p_0,p_1,\cdots, p_m$ which is also interesting
in $Q_{i+1}$. (Note that $Event(j)$ and $Event(j')$ are excluding
for any $j\neq j'$.) Then $j$ is the number of searching steps
that will be performed in $Q_i$. We overlook the case $j=0$ since
it contributes nothing to the expected value of $j$.

Since each interesting square has at least two non-empty quarters,
there are at least $j+1$ non-empty quarters hung off the subpath
$p_{m-j},\cdots,p_m$. $Event(j)$ occurs only if one (with
probability $Pr_1$) or zero (with probability $Pr_0$) quarters
among these $\geq j+1$ quarters is still non-empty in $Q_{i+1}$.
Otherwise, the LCA of the two non-empty quarters in $Q_i$ will be
interesting in $Q_{i+1}$. So $Pr(j)\leq Pr_1+Pr_0\leq
\frac{j+1}{2^{j+1}}+\frac{1}{2^{j+1}}.$ (E.g., consider
$p_{m-j}=p_{0}$ in Fig.~\ref{Fig-skip-1}. Note that this is not a
tight upper bound.) The expected value of $j$ is then

\begin{equation}
E(j)=\sum_1^m jPr(j)\leq\sum_1^m
j(\frac{j+1}{2^{j+1}}+\frac{1}{2^{j+1}}) =\frac{1}{2}\sum_1^m
\frac{j^2}{2^{j}}+\sum_1^m\frac{j}{2^{j}}\approx \frac{1}{2}\times
6.0+2.0=5.0.
\end{equation}

Consider an example that each $p_j$ has exact two non-empty
quarters, and the non-empty quarter of $p_j$ that does not contain
$p_{j+1}$ contains exact one point $x_j$. (For the two non-empty
quarters of $p_m$, we let each of them contain exact one point,
and choose any one as $x_m$.) $Event(j)$ happens iff $x_j$ is
selected to $S_{i+1}$ and another point among the rest $m-j+1$
points contained in $p_j$ is also selected. So
$Pr(j)=\frac{1}{2}\cdot\frac{j+1}{2^{j+1}}$. (E.g., consider
$p_{m-j}=p_{1}$ in Fig.~\ref{Fig-skip-1}.) The expected value of
$j$ is then

\begin{equation} E(j)=\sum_1^m jPr(j)=\frac{1}{4}(\sum_1^m
\frac{j^2}{2^{j}}+\sum_1^m\frac{j}{2^{j}})\approx
\frac{1}{4}(6.0+2.0)=2.0.
\end{equation}

Therefore in the worst case, the expectation of $j$ is between 2
and 5. (See the appendix for out computation of the progressions in (1) and
(2).)
\end{proof}

To insert a point $x$ into the structure, we perform the above
point location search which finds $p_{i,end}$ within all the
$Q_i$'s, flip coins to find out which $S_i$'s $x$ belongs to, then
for each $S_i$ containing $x$, insert $x$ into $p_{i,end}$ in
$Q_i$ as described in Section~\ref{compressed}. Note that by
flipping coins we may create one or more new non-empty subsets
$S_{l+1},\cdots$ which contains only $x$, and we shall
consequently create the new compressed quadtrees $Q_{l+1},\cdots$
containing only $x$ and add them into our data structure. Deleting
a point $x$ is similar. We do a search first to find $p_{i,end}$
in all $Q_i$'s. Then for each $Q_i$ that contains $x$, delete $x$
from $p_{i,end}$ in $Q_i$ as described in
Section~\ref{compressed}, and remove $Q_i$ from our data structure
if $Q_i$ becomes empty.

{\theorem Searching, inserting or deleting
any point in a randomized $d$-dimensional
skip quadtree of $n$ points takes expected $O(\log n)$ time.}

\begin{proof} The expected number of non-empty subsets is obviously
$O(\log n)$. Therefore by Lemma~\ref{25}, the expected searching
time for any point is $O(1)$ per level, or $O(\log n)$ overall.
For any point to be inserted or
deleted, the expected number of non-empty subsets containing this
point is obviously $O(1)$. Therefore the searching time dominates
the time of insertion and deletion.
\end{proof}

In addition, note that the expected space usage for a skip quadtree
is $O(n)$, since the expected size of the compressed quadtrees in the
levels of the skip quadtree forms a geometrically decreasing sum that
is $O(n)$.

\section{The Deterministic Skip Quadtree}\label{deterministic}

\begin{figure}[hbt]
\vspace*{-10pt}
\begin{center}
\includegraphics[scale=0.6]{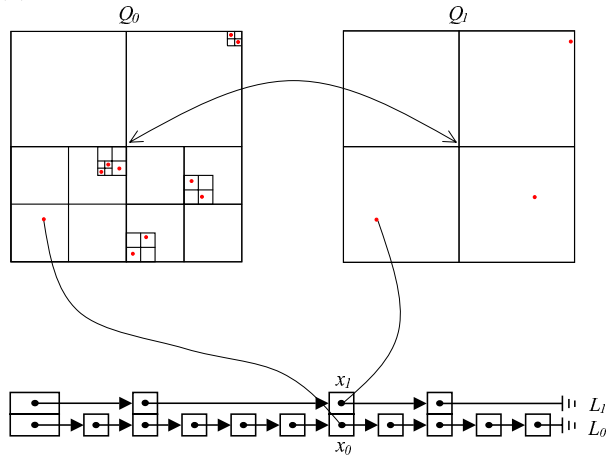}
\end{center}
\vspace*{-10pt} \caption{A deterministic skip quadtree guided by a
deterministic 1-2-3 skip list.}\label{Fig-skip-2}
\end{figure}

In the deterministic version of the skip quadtree data structure,
we again maintain a sequence of subsets $S_i$ of the input points
$S$ with $S_0=S$ and build a compressed quadtree $Q_i$ for each
$S_i$. However, in the deterministic case,
we make each $Q_i$ an ordered tree and sample $S_i$
from $S_{i-1}$ in a different way. We can order the $2^d$ quarters
of each $d$-dimensional square (e.g., by the I, II, III, IV quadrants in $\R^2$
as in Fig.~\ref{Fig-skip-0} or by the
lexical order of the $d$-dimensional coordinates in high dimensions), and call
a compressed quadtree obeying such order an \emph{ordered
compressed quadtree}. Then we build an ordered compressed quadtree
$Q_0$ for $S_0=S$ and let $L_0=L$ be the ordered list of $S_0$ in
$Q_0$ from left to right. Then we make a skip list $\mathcal{L}$
for $L$ with $L_i$ being the $i$-th level of the skip list. Let
$S_i$ be the subset of $S$ that corresponds to the $i$-th level
$L_i$ of $\mathcal{L}$, and build an ordered compressed quadtree
$Q_i$ for each $S_i$. Let $x_i$ be the copy of $x$ at level $i$ in
$\mathcal{L}$ and $p_i(x)$ be the smallest interesting square in
$Q_i$ that contains $x$. Then, in addition to the pointers in
Sec.~\ref{randomized}, we put a bi-directed pointer between $x_i$
and $p_i(x)$ for each $x\in S$. (See Fig.~\ref{Fig-skip-2}.)

{\lemma The order of $S_i$ in $Q_i$ is $L_i$.}

\begin{proof}
Noting that an interesting square in $Q_i$ is also an interesting
square in $Q_{i-1}$, the LCA of two points $x$ and $y$ in $Q_i$ is
also a CA of them in $Q_{i-1}$. Therefore the order of $S_i$ in
$Q_i$ is a subsequence of the order of $S_{i-1}$ in $Q_{i-1}$. By
induction, the order of $S_i$ in $Q_i$ is $L_i$, given that the
order of $S_0$ in $Q_0$ is $L_0$. \end{proof}

The skip list $\mathcal{L}$ is implemented as a deterministic
1-2-3 skip list in~\cite{Munro&PS02}, which maintains the property
that between any two adjacent columns at level $i$ there are 1, 2
or 3 columns of height $i-1$ (Fig.~\ref{Fig-skip-2}). There are
$O(\log n)$ levels of the 1-2-3 skip list, so searching takes
$O(\log n)$ time. Insertion and deletion can be done by a search
plus $O(1)$ promotions or demotions at each level along the
searching path.

We also binarize $Q_0$ by adding $d-1$ levels of dummy nodes, one
level per dimension, between each interesting square and its $2^d$
quarters. Then we independently maintain a total order (the
in-order in the binary $Q_0$) for the set of interesting squares,
dummy nodes and points in $Q_0$. The order is maintained as
in~\cite{D&Sleator87order} which supports the following
operations: 1) insert $x$ before or after some $y$; 2) delete $x$;
and 3) compare the order of two arbitrary $x$ and $y$. All
operations can be done in deterministic worst case constant time.
These operations give a total order out from a linked list, which
is necessary for us to search in $\mathcal{L}$. Because of the
binarization of $Q_0$ and the inclusion of all internal nodes of
the binarized $Q_0$ in our total order,
when we insert a point $x$ into $Q_0$, we get a $y$ (parent of $x$
in the binary tree) before or after the insertion point of $x$, so that
we can accordingly insert $x$ into our total order.

\ifFull
\else

In the full version, we give details for insertion and deletion in a
deterministic skip quadtree (searching is the same as in the
randomized version), proving the following:
\fi

{\theorem Search, insertion and deletion in a deterministic $d$-dimensional skip
quadtree of $n$ points take worst case $O(\log n)$ time.}

Likewise, the space complexity of a deterministic skip quadtree is
$O(n)$.

\section{Approximate Range Queries}
In this section, we describe how to use a skip quadtree to perform
approximate range queries, which are directed at reporting the
points in $S$ that belong to a query region (which can be an
arbitrary convex shape having $O(1)$ description complexity). For
simplicity of expression, however, we assume here that the query
region is a hyper-sphere. We describe in the full version how to
extend our approach to arbitrary convex ranges with constant
description complexity. We use $x,y$ for points in the data set
and $u,v$ for arbitrary points (locations) in $\R^d$. An
approximate range query with error $\epsilon>0$ is a triple
$(v,r,\epsilon)$ that reports all points $x$ with $d(v,x)\leq r$
but also some arbitrary points $y$ with $r<d(v,y)\leq
(1+\epsilon)r$. That is, the query region $R$ is a (hyper-) sphere
with center $v$ and radius $r$, and the permissible error range
$A$ is a (hyper-) annulus of thickness $\epsilon r$ around $R$.

Suppose we have a space partition tree $T$ with depth $D_T$ where
each tree node is associated with a region in $\R^d$. Given a
query $(v,r,\epsilon)$ with region $R$ and annulus $A$, we call a
node $p\in T$ an \emph{in}, \emph{out}, or \emph{stabbing} node if
the $\R^d$ region associated with $p$ is contained in $R\cup A$,
has no intersection with $R$, or intersects both $R$ and
$\overline{R\cup A}$. Let $S$ be the set of stabbing nodes in $T$
whose child nodes are not stabbing. Then the query can be answered
in $O(|S|D_T+k)$ time, with $k$ being the output size. Previously
studied space partition trees, such as BBD
trees~\cite{am-annqf-93,am-ars-00,amnsw-oaann-98} and BAR
trees~\cite{d-bart-99,dgk-bartc-01}, have an upper-bound on $|S|$
of $O(\epsilon ^{1-d})$ and $D_T$ of $O(\log n)$, which is
optimal~\cite{am-ars-00}. The ratio of the unit volume of $A$,
$(\epsilon r)^d$, to the lower bound of volume of $A$ that is
covered by any stabbing node is called the packing function
$\rho(n)$ of $T$, and is often used to bound $|S|$. A constant
$\rho(n)$ immediately results in the optimal $|S|=O(\epsilon
^{1-d})$ for convex query regions, in which case the total volume
of $A$ is $O(\epsilon r^d)$.

Next we'll show that a skip quadtree data structure answers an
approximate range query in $O(\epsilon ^{1-d}\log n+k)$ time. This
matches the previous results (e.g.,
in~\cite{d-bart-99,dgk-bartc-01}), but skip quadtrees are fully
dynamic and significantly simpler than BBD trees and BAR trees.

\subsection{Query Algorithm and Analysis}
Given a skip quadtree $Q_0,Q_1,\cdots,Q_l$ and an approximate
range query $(v,r,\epsilon)$ with region $R$ and annulus $A$, we
define a \emph{critical square} $p\in Q_i$ as a stabbing node of
$Q_i$ whose child nodes are either not stabbing, or still stabbing
but cover less volume of $R$ than $p$ does. If we know the set
$C=C_0$ of critical squares in $Q_0$, then we can answer the query
by simply reporting, for each $p\in C$, the points inside every 
in-node child square of $p$ in $Q_0$. We now show that the size of
$C$ is $O(\epsilon ^{1-d})$ (due to the obvious constant $\rho$ of
quadtrees).

{\lemma The number of critical squares in $Q_0$ is $O(\epsilon
^{1-d})$.\label{pack}}

\begin{proof} Consider the inclusion tree $T$ for the critical
squares in $C$ (that is, square $p$ is an ancestor of square $q$
in $T$ iff $p\supseteq q$ ). We call a critical square a branching
node if it has at least two children in $T$, or a non-branching
node otherwise. A non-branching node either is a leaf of $T$, or
covers more volume of $R$ than its only child node in $T$ does, by
the definition of critical squares. Note that if two quadtree
squares cover different areas (not necessarily disjoint) of $R$,
then they must cover different areas of $A$. Therefore for each
non-branching node $p\in T$, there is a unique area of $A$ covered
by $p$ but not by any other non-branching nodes of $T$. The volume
of this area is clearly $\Omega(1)$ of the unit volume $(\epsilon
r)^d$ since $p$ is a hypercube. Thus the total number of
non-branching nodes in $T$ is $O(\epsilon ^{1-d})$ since the total
volume of $A$ is $O(\epsilon r^d)$. So $|C|=|T|$ is also
$O(\epsilon ^{1-d})$.
\end{proof}

Next we complete our approximate range query algorithm by showing
how to find all critical squares in each $Q_i$. The critical
squares in $Q_i$ actually partition the stabbing nodes of $Q_i$
into equivalence classes such that the stabbing nodes in the same
class cover the same area of $R$ and $A$. Each class corresponds
to a path (could be a single node) in $Q_i$ with the tail (lowest)
being a critical square and the head (if not the root of $Q_i$)
being the child of a critical square. For each such path we
color the head red and the tail green (a node could be double
colored). The set of green nodes (critical squares) is $C_i$.

Assume we have the above coloring done in $Q_{i+1}$. We copy the
colors to $Q_i$ and call the corresponding nodes $p\in Q_i$
\emph{initially green} or \emph{initially red} accordingly. Then
from each initially green node we DFS search down the subtree of
$Q_i$ for critical squares. In addition to turning back at each
stabbing node with no stabbing child, we also turning back at each
initially red node. During the search, we color (newly) green or
red to the nodes we find according to how the colors are defined
above. After we've done the search from all initially green nodes,
we erase the initial colors and keep only the new ones.

{\lemma The above algorithm correctly finds the set $C_i$ of all
critical squares in $Q_i$ in $O(|C_{i+1}|)$ time, so that finds
$C=C_0$ in $O(|C|\log n)$ time, which is the expected running time
for randomized skip quadtrees or the worst case running time for
deterministic skip quadtrees.\label{T}}
%
%

\begin{proof}
Correctness. If a stabbing node $p$ has its closest initial red
ancestor $p'$ lower than its closest initial green ancestor $p''$,
then it will be missed since when searching from $p''$, $p$ will
be blocked by $p'$. Let $p'$ and $p''$ be a pair of initially red
and green nodes which correspond to the head and tail of a path of
equivalent stabbing squares in $Q_{i+1}$, and $p$ be a stabbing
square in $Q_i$ that is a descendant of $p'$. Note that, if $p$
covers less area of $R$ than $p'$ and $p''$ do, then $p$ is also a
descendant of $p''$; otherwise if $p$ covers the same area of $R$
as $p'$ and $p''$ do but $p$ is not a descendant of $p''$, then
$p$ is not critical because $p''$ is now contained in $p$.
Therefore we won't miss any critical squares.

Running time.
\ifFull
\else By the same arguments as in Lemma~\ref{25},
\fi
the DFS search downward each initially green node $p$ has
constant depth, because within constant steps we'll meet a square
$q$ which is a child node of $p$ in $Q_{i+1}$. $q$ is either not
stabbing or colored initially red so we'll go back.
\end{proof}

Following Lemma~\ref{pack},\ref{T} and the algorithm description,
we immediately get

{\theorem We can answer an approximate range query
$(v,r,\epsilon)$ in $O(\epsilon ^{1-d}\log n+k)$ time with $k$
being the output size, which is the expected running time for
randomized skip quadtrees or the worst case running time for
deterministic skip quadtrees.}

{\footnotesize
\bibliographystyle{abbrv}
\bibliography{skip,extra,geom}
}

\clearpage
\begin{appendix}
\section{Appendix}
If $f(x)\geq 0$ is a monotone decreasing function for $x\geq i$,
then the progression of $f(x)$ can be approximated by its integral
as following:
$$\sum_{x=1}^{i-1}f(x)+\int_{x=i}^{\infty}f(x)dx\leq
\sum_{x=1}^{\infty}f(x)\leq
\sum_{x=1}^{i}f(x)+\int_{x=i}^{\infty}f(x)dx.$$ By

$$\int_{x=i}^{\infty}\frac{x^2}{2^x}dx=\frac{1}{2^x\ln^{3}2}[(\ln2\cdot x)^2+2\ln2\cdot
x+2]\mid_{x=i}$$ and

$$\int_{x=i}^{\infty}\frac{x}{2^x}dx=\frac{1}{2^x\ln^{2}2}(\ln2\cdot
x+1)\mid_{x=i}$$ we get (taking $i=12$)
$$\sum_{1}^{\infty}\frac{x^2}{2^{x}}\leq 6.01$$ and (taking
$i=10$)
$$\sum_{1}^{\infty}\frac{x}{2^{x}}\leq 2.005.$$

\end{appendix}

\end{document}